# Simulation of an Optimum Multilevel Dynamic Round Robin Scheduling Algorithm

Neetu Goel
Research Scholar
Teerthanker Mahaveer University,
Moradabad, India

R.B. Garg, Ph.D
Professor
Tecnia Institute of Advanced Studies,
Rohini, Delhi

## ABSTRACT

CPU scheduling has valiant effect on resource utilization as well as overall quality of the system. Round Robin algorithm performs optimally in timeshared systems, but it performs more number of context switches, larger waiting time and larger response time. The devised tool "OMDRRS" was utilized to simulate the four algorithms (FCFS, SJF, ROUND ROBIN &  Proposed Dynamic Round Robin Algorithm) utilizing either manual entered process with burst time as well as system generated processes with randomly generated burst time. In order to simulate the behavior of various CPU scheduling algorithms and to improve Round Robin scheduling algorithm using dynamic time slice concept,  in this paper we produce the implementation of new CPU scheduling algorithm called An *Optimum Multilevel Dynamic Round Robin Scheduling (OMDRRS),* which calculates intelligent time slice and warps after every round of execution. The results display the robustness of this software, especially for academic, research and experimental use, as well as proving the desirability and efficiency of the probabilistic algorithm over the other existing techniques and it is observed that this OMDRRS projects  good performance as compared to the other existing CPU scheduling algorithms.

## 1. INTRODUCTION

The scheduling simulator illustrates the behavior of scheduling algorithms against a simulated mix of process loads. It is a framework that lets you to swiftly and easily devise and collect metrics for custom CPU scheduling strategies. There are a number of such algorithms with each having its respective advantages and drawbacks. In order to calculate the comparative and competitive advantages and disadvantages of these algorithms, the algorithm requires to be simulated and their performance indices studied and utilized for better capturing of operating system principles. Some of these algorithms would reflect  promising results in terms of ease of implementation but perform poorly in terms of turnaround time, waiting time, context switch and vice versa. In Round Robin (RR) every process has equal priority and is provided a time quantum or time slice after which the process is preempted. Although Round Robin displays improved response time and utilizes shared resources effectively, its limitations are larger waiting time, undesirable overhead and larger turnaround time for processes with inconstant  CPU bursts due to use of static time quantum. This motivates us to implement Round Robin algorithm with dynamic burst time concept.

To properly illustrate the functionality of various CPU scheduling algorithms and improvement of Round Robin scheduling using dynamic time slice concept called "Dynamic Round Robin" was depicted  using VB6.0 and the results of all algorithms were collected and analyzed with the help of Turnaround time, waiting time, Context Switch & Gantt Chart.

### 1.1 Organization of the Paper

This paper is sliced into five sections. *Section 1* projects a brief introduction on the various aspects of the scheduling algorithms, the approach to the current paper and the motivational factors leading to this improvement. *Section 2* reflects an overview of some of the simulators that are available and their respective drawbacks. A brief overview, characteristics and flaws of some of the existing process scheduling algorithms are discussed in *Section 3*. *Section 4* describes the datasets, design issues, mode of operation, and the details of implementation of the simulator on the algorithms. Results also show the comparative performance of the four algorithms in this section. Conclusion is presented in *Section 5* followed up by the references used.

## 2. EXISTING SIMULATORS

Process Scheduling Simulator[5] is a java-based web application that implements FCFS, SJF, Priority SJF and Round Robin. It requires a high-speed internet connection to load the applet, and also requires that Java software to be either installed or updated. Each input in the system is visualized by its arrival time, CPU burst and I/O bursts. It claims to be very efficient but a sample run divulged that it is very slow. Another simulator "CPU Scheduling Simulator (CPUSS)" [6], it is a framework that permits users to swiftly and easily devise and collects metrics for custom CPU scheduling strategies including FCFS, Round Robin, SJF, Priority First, and SJF with Priority Elevation rule. The long list of the capabilities it can possess proves it too complex and complicated for simple academic demonstrations and utilize by non-computer geeks such as greenhorn students that are just taking their first course in Computer Science. Above all, it flows in the windows-DOS environment which is characterized by no lucrative user interface and hence, lacks user-friendliness. A project that is very close to our work is a simulator presented by (Padberg, 2003)[7]. However, this simulator was devised for a software project scheduling rather than CPU process scheduling, hence impertinent for our consideration in this study. MOSS[8], Modern Operating Systems Simulators, it is a bible of Java-based simulation programs which illustrate key operating system concepts portrayed in a textbook by Tanenbaum (2001) for university students utilizing the text. This does not suit in to independent software that can be utilized freely without any such constraint. The best simulator we could find, so far, during our survey of previous related work was presented by (Cardella, 2002)[9]. It was developed in Visual Basic 6.0 and implemented the Round Robin as a non-preemptive scheduling algorithm. It uses Average Completion Time (ACT) and Average Turn-around Times (ATT) as the criteria for performance evaluation. However, it is not as robust as ours in the sense that we implemented a Dynamic Round Robin algorithm in addition to FCFS, SJF and ROUND ROBIN algorithms. Our major objective is to simulate the behavior of various CPU scheduling algorithms and to improve Round Robin scheduling algorithm using dynamic time slice concept, called *Dynamic Round Robin*, which calculates intelligent time slice and changes after every round of execution.

## 3. CONVENTIONAL PROCESS SCHEDULING ALGORITHMS

### 3.1 First Come First Serve

The ultimate intuitive and down to earth technique is to permit the first process submitted to flow first. This technique is called as first-come, first-served (FCFS) scheduling. In effect, processes are inserted into the tail of a queue when they are submitted. The next process is picked from the head of the queue when each finishes running.

### Characteristics

- The drawback of prioritization does permit every process to eventually fulfill, hence no starvation.
- Turnaround time, waiting time and response time is at the acme.
- One, process with longest burst time can monopolize CPU, even if other process burst time is too short. Hence throughput is low [12].

### 3.2 Non preempted Shortest Job First

The process is sanctioned to the CPU which has minimum burst time. A scheduler arranges the processes with the minimum burst time in head of the queue and longest burst time in tail of the queue. This strives advanced knowledge or estimations about the time required for a process to complete [1]. This algorithm is devised for maximum throughput in most scenarios.

### Characteristics

- The real difficulty with the SJF algorithm is, to know the length of the next CPU request.
- SJF minimizes the average waiting time [12] because it services small processes before it services large ones. While it minimizes average wait time, it may penalize processes with high service time requests.

### 3.3 Round Robin

The Round Robin (RR) scheduling algorithm allocates a small unit of time, called time slice or quantum time. The ready processes are kept in a queue. The scheduler goes in the order of this queue, allocating the CPU to each process for a time interval of assigned quantum. New processes are added to the tail of the queue [13].

### Characteristics

- Setting the quantum too short originate too many context switches and lower the CPU efficiency.
- Setting the quantum too long may cause poor response time and fairly nearby the FCFS.
- Because of high waiting times, deadlines are rarely met in a pure RR system.

### 3.4 Priority Scheduling

The operating system provides a fixed priority rank to each process. Lower priority processes get interrupted by incoming higher priority processes.

### Characteristics

- Starvation can happen to the low priority process.
- The waiting time gradually increases for the equal priority processes [14].
- Higher priority processes have smaller waiting time and response time.

## 4. DESIGN, IMPLEMENTATION AND DATA DESCRIPTION

### 4.1 Our Proposed Algorithm

In our algorithm, combines the fundamental principles of various scheduling algorithms as well as the dynamically Time Slice (DTS) concept based on priority, shortest CPU burst time. Main steps are:

**Step 1:** Shuffle the processes in ascending order in the ready queue such that the head of the ready queue contains the lowest burst time.

**Step 2:** If one or more process has equal burst time then

{

Allocate the CPU to the processes according to First Come basis.

}

**Step 3:** Assign the time quantum and apply for each process say TQ=k.

**Step 4:** IF (burst time of the process < TQ)

{

Allocate the CPU to that process till it terminates.

}

ELSE IF (Remaining burst time of the process < TQ/F)

{

Allocate the CPU again to that process till it terminates.

}
ELSE
{

(i) The process will occupy the CPU till the time quantum and it is added to the ready queue in ascending order for the next round of execution.

(ii) TQ= TQ * F

(iii) TQ= K

(iv) Goto Step 3

}

## 4.2 Software Design

The simulator *OMDRRS* was designed and developed using the Microsoft Visual Basic 6.0 Professional Edition's Integrated Developed Environment (IDE). The input data were created either as Manual Process Entry with burst time as well as Automatic Process Generator with randomly generated burst time. In Automatic Process Generator system, it fetches all the active processes with randomly generated burst time while in the manual entered process user entered the burst time as per their requirement. Based on the selected input type: *1) Manual Process Entry 2) Automatic Process Generator* and the scheduling algorithm FCFS, SJF, Round Robin and the Dynamic Round Robin algorithm were computed and display the ATT(Average Turnaround Time), AWT(Average Waiting Time), CS(Context Switch) and Gantt Chart were automatically generated & displayed at runtime. The result of each algorithm is also displayed on a window for the user to view. The *OMDRRS* software was designed as a simple, light-weight system for academic as well as the research purpose for the simulation of the behavior of FCFS, SJF, Round Robin and Dynamic Round Robin scheduling algorithms. Quality is further strengthened with the fact that the entire software does not exceed 4MB in size. The user interfaces are simple, concise, unambiguous and easy to use but replete with only the relevant information. The input of burst time is re-useable for comparing with all other algorithms. The innovative Dynamic algorithm is well implemented and its mode of operation was clearly shown and presented in the simulator.

## 4.3 Implementation

The software was implemented to simulate the procedure of FCFS, SJF, Round Robin and Improving of Round Robin scheduling algorithm. These algorithms were implemented in order to establish a valid premise for effective comparison. The simulator takes process IDs as integer, randomly generated burst times and their respective positions in terms of their order like in a virtual queue. For simplicity, the simulator was built on three major assumptions:

- The scheduling policy of FCFS & SJF are non-preemptive,
- The quantum time of Round Robin and Dynamic Algorithm are generated randomly.
- All the Processes arrive at the same time.

The *Automatic Process Generator* simulator was run on different datasets depending on how many applications were activate in the existing system with randomly produce burst time that have been positioned in queue for the process arrival scenarios in the system. This was done to determine, as part of the experiment, whether the location of a process in a queue will affect the results of the entire simulation algorithm. The simulation was run several times to ensure fairness to all datasets and presented for each algorithm using Average Turn-around Time, Average Waiting Time, Context Switch and Gantt chart as the performance evaluation indices.

The *Manual Process Entry* simulator tool was run on different datasets depending on user requirements that have been positioned in queue for the process arrival scenarios in the system. This was done to determine, as part of the experiment, whether the location of a process in a queue will affect the results of the all simulation algorithm. The simulation was run several times to ensure fairness to all datasets and presented for each algorithm using Average Turn-around Time, Average Waiting Time, Context Switch and Gantt chart as the performance evaluation indices.

## 4.4 Description of Data

(Table 1) & (Table 2) shows the datasets representing processes that are identified by their IDs, with their randomly generated burst times and the output of FCFS, SJF, RR, Dynamic RR system generated Turn Around Time (TAT), Waiting Time (WT) according to the Automatic Process Generator & Manual Process Entry. The different arrangement of the jobs was intended to significant the different real-world scenario, jobs can take with different estimated burst times on the waiting queue. The number of processes can be extended to any length as desired. For demonstration purpose, a maximum of 10 jobs in the Automatic Process Generator (APG) & 5 jobs in the Manual Process Entry (MPE) were taken and reported in this paper. It was tested with 20 and 30 jobs during testing. However, the maximum attainable number of jobs was not determined because it totally depends on the Memory size.

**Table 1: Assume ten processes arrived at time =0 with randomly generated burst time and simulator generated automatically TAT, WT as per scheduling policy**

| APG | | FCFS | | SJF(Non Preemptive) | | RR | | Dynamic RR | |
|---|---|---|---|---|---|---|---|---|---|
| PID | BT | TAT | WT | TAT | WT | TAT | WT | TAT | WT |
| P1 | 16 | 16 | 0 | 74 | 58 | 134 | 118 | 94 | 78 |
| P2 | 13 | 29 | 16 | 43 | 30 | 106 | 93 | 73 | 60 |
| P3 | 15 | 44 | 29 | 58 | 43 | 111 | 96 | 83 | 68 |
| P4 | 10 | 54 | 44 | 18 | 8 | 70 | 60 | 58 | 48 |
| P5 | 12 | 66 | 54 | 30 | 18 | 113 | 101 | 65 | 53 |
| P6 | 22 | 88 | 66 | 96 | 74 | 156 | 134 | 141 | 119 |
| P7 | 8 | 96 | 88 | 8 | 0 | 83 | 75 | 53 | 45 |
| P8 | 24 | 120 | 96 | 120 | 96 | 160 | 136 | 160 | 136 |
| P9 | 26 | 146 | 120 | 171 | 145 | 171 | 145 | 171 | 145 |
| P10 | 25 | 171 | 146 | 145 | 120 | 170 | 145 | 165 | 140 |

**Table 2: Assume five processes arrived at time =0 with user entered the burst time and simulator generated automatically TAT, WT as per scheduling policy**

| MPE | | FCFS | | SJF(Non Preemptive) | | RR | | Dynamic RR | |
|---|---|---|---|---|---|---|---|---|---|
| PID | BT | TAT | WT | TAT | WT | TAT | WT | TAT | WT |
| P1 | 15 | 15 | 0 | 26 | 11 | 50 | 35 | 38 | 23 |
| P2 | 20 | 35 | 15 | 46 | 26 | 64 | 44 | 52 | 32 |
| P3 | 7 | 42 | 35 | 11 | 4 | 41 | 34 | 11 | 4 |
| P4 | 30 | 72 | 42 | 76 | 46 | 76 | 46 | 76 | 46 |
| P5 | 4 | 76 | 72 | 4 | 0 | 28 | 24 | 4 | 0 |

From the above comparisons, it is apparent that the dynamic time quantum approach is more effective then the fixed time quantum approach in terms of turnaround time, waiting time and context switch.

## 4.5 Experimental Computing Environment

We conducted a simulation-based experimental study that runs on a laboratory Personal Computer with the Service Park 3 update of Windows XP Professional Edition version 2002. The processor is based on Intel(R) Core (TM)2 Duo CPU with a speed of 2.93 GHz and a RAM size of 1.96 GB. Hence the ATT, AWT, CS and Gantt chart used as criteria for performance evaluation to validate the results. (Table 3) portray the assessment between FCFS, SJF, Round Robin and the proposed algorithm based on data given in the (*Table 1*). (Table 4) depicts the assessment between FCFS, SJF, Round Robin and the proposed algorithm based on (*Table 2*).

**Table 3: Si**mulator generated automatically Average Turnaround Time(ATT), Average Waiting Time(AWT), Context Switch (CS)

| SCH. CRITERIA | FCFS | SJF | ROUND ROBIN ALGO | DYNAMIC RR |
|---|---|---|---|---|
| CONTEXT SWITCH | 10 | 10 | 38 | 27 |
| TURNAROUND TIME | 83 | 76.3 | 127.4 | 106.3 |
| WAITING TIME | 65.9 | 59.2 | 110.3 | 89.2 |

**Table 4: Si**mulator generated automatically Average Turnaround Time(ATT). Average Waiting Time(AWT), Context Switch (CS)

| SCH. CRITERIA | FCFS | SJF | ROUND ROBIN ALGO | DYNAMIC RR |
|---|---|---|---|---|
| CONTEXT SWITCH | 5 | 5 | 15 | 10 |
| TURNAROUND TIME | 48 | 32.6 | 51.8 | 36.2 |
| WAITING TIME | 32.8 | 17.4 | 36.6 | 21 |

## 4.6 RESULTS AND DISCUSSIONS

This section depicts the graphically representation of comparison of the proposed algorithm with the existing algorithm based on the average turnaround time, average waiting time and context switches. Results for the Automatic Process Generator using 10 processes using four scheduling algorithm as well as in the Manual Process Entry using 5 processes using four scheduling algorithm. (Fig. 1) shows the bar graph of Automatic Process Generator based on (Table 3). (Fig. 2) shows the bar graph of Manual Process Entry based on (Table 4*).*

**Fig. 1: Bar graph of Simulator generated automatically Average Turnaround Time (ATT), Average Waiting Time(AWT), Context Switch (CS) of ten processes**

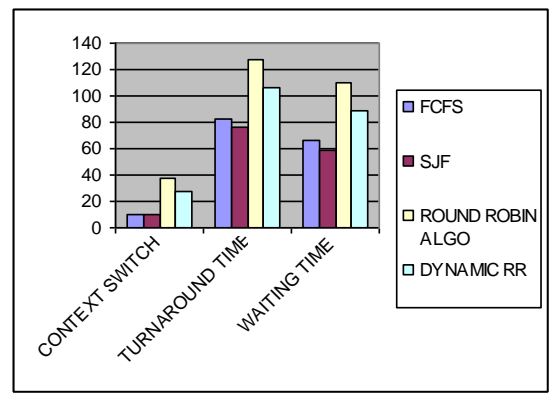

**Fig. 2:** Bar graph of Simulator generated automatically Average Turnaround Time (ATT), Average Waiting Time(AWT), Context Switch (CS) of five processes

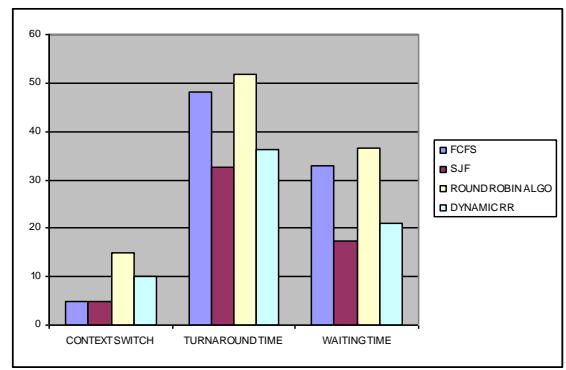

## 5. CONCLUSION

Simulator (OMDRRS) has been developed. OMDRRS has presented a light-weight simulator which depicts First Come First Serve, Shortest Job First, Round Robin and improvement of Round Robin Scheduling. Simulator (OMDRRS) Software, comparing the efficiency and performance in terms of Average Turn-around Time, Average Waiting Time, Context Switch and Gantt chart. Ready queue is maintained as a FIFO queue to implement all the major algorithms. Processes are selected from the head of the ready queue. A preempted process is linked at the tail of the ready queue. Dynamic Round Robin Algorithm has proven, on an average, to be very fair to the process to be selected from the ready queue, and quick in terms of execution time. Each process, having fair chances, is scheduled by random sampling from among waiting processes in the ready queue. It is analyzed that the Dynamic Scheduling algorithm is superior as it has less waiting response time, usually less pre-emption and context switching thereby reducing the overhead and saving of memory space.